# What factors have caused Japanese prefectures to attract a larger population influx?

Keisuke Kokubun[1]

1 Economic Research Institute, Japan Society for the Promotion of Machine Industry

## Abstract

Regional promotion and centralized correction in Tokyo have long been the goals of the Government of Japan. Furthermore, in the wake of the recent new coronavirus (COVID-19) epidemic, the momentum for rural migration is increasing, to prevent the risk of infection with the help of penetration of remote work. However, there is not enough debate about what kind of land will attract the population. Therefore, in this paper, we will consider this problem by performing correlation analysis and multiple regression analysis with the inflow rate and the excess inflow rate of the population as the dependent variables, using recent government statistics for each prefecture. As a result of the analysis, in addition to economic factor variables, variables of climatic, amenity, and human factors correlated with the inflow rate, and it was shown that the model has the greatest explanatory power when multiple factors were used in addition to specific factors. Therefore, local prefectures are required to take regional promotion measures focusing on not only economic factors but also multifaceted factors to attract the outside population.

## Introduction

Japan has a heavily concentrated population in Tokyo, the capital of Japan, and its problems have been discussed concerning efficiency, urban traffic congestion, housing shortages, environmental pollution, disaster risk, and rural depopulation (Hirata, 2019; Porter, 1990). Therefore, the government has aimed to correct the overconcentration in Tokyo and promote the development of local areas through financial support, employment creation, etc. by implementing the national comprehensive development plan. However, the effect is limited, and there are no signs of improvement in recent years. In response to this situation, the recent corona pandemic has forced people living in cities to reassess the value of local prefectures with a low risk of infection. Also, the penetration of remote work, which is a work style that can be used anywhere, has made it possible for them to move to rural areas. These changes are expected to boost population migration from urban to rural areas.

If the relative attractiveness of Tokyo is diminishing, the corona pandemic may be an opportunity to correct the overconcentration in Tokyo and promote regional development. However, if economic factors are the only factors that influence migration, the scale of migration may also be limited, as rural areas with scarce economic resources have limited ability to attract population. In this regard, salvation is that not a few previous studies have shown that in addition to economic factors, human factors, amenity factors, and climatic factors influence population migration. For example, if people can be attracted by utilizing human networks, there are expectations for regional development even in prefectures where the jobs of industries that can pay high salaries are small. However, as far as the author knows, there are no studies that have analyzed how better the population migration can be explained by combining these different factors compared to the case where they act alone. To consider migration and its factors in the age of With-Corona, it is necessary to correctly understand not only economic factors but also the effects of combined various factors. Therefore, this paper will tackle this theme by hierarchical multiple regression analysis using the population inflow rate and inflow excess rate as dependent variables and various factors as independent variables using the data set of 47 prefectures in Japan obtained from the government statistics.

## Examination of previous research and presentation of hypothesis

*Economic factors*

Economic factors are the most representative and indispensable element in classical economics. Therefore, many studies have revealed that local employment and income/wage differences account for inter-regional migration (Arntz, 2010; Asada, 1996; Attanasio & Padoa-Schioppa, 1991; Faggian et al. 2007; Ferguson et al. 2007; Higuchi, 1991; Itoh, 2006; Lee, 2012; Ohta & Ohkusa, 1996; Scott 2010; Storper & Scott 2009; Tachi 1963; Tanioka, 2001; Toyota, 2013; Vakulenko, 2016; Watanabe, 1994; Zhang et al., 2016). Some variables harm migration. For instance, a study suggests that people tend to leave metropolitan areas with high costs of living (Herzog & Schlottmann, 1986). From the above, the following hypothesis is derived.

H1: Economic factors correlate with the population inflow rate.

*Climatic factors*

Few studies have examined the effects of climate on migration. It has been pointed out that in the United States, after a certain period, the migration from the high-income and cold northern part to the low-income and warm southern part became predominant (Graves, 1980). A study of Japan also showed that after the period of a high economic miracle, there was a gradual change from excess inflow to cold regions to excess inflow to warm regions (Ito, 2006). Furthermore, although the study covers a more limited area, the study by Tomioka & Sasaki (2003) shows that there is a positive correlation between the minimum temperature and the population inflow rate. Therefore, the following hypothesis is derived.

H2: Climatic factors correlate with population influx.

*Amenity factors*

It has been pointed out that the uneven distribution of amenities (living environment), including road infrastructure development, maybe causing the overconcentration of the population (Hirata et al., 2019). However, few empirical studies incorporate amenity factors into the model compared to economic factors such as income and employment opportunities (Liu & Shen, 2014; Zhang et al., 2016). Also, there are negative suggestions on the effects of amenity factors. For example, studies have shown that well-educated working-age populations move with employment opportunities as an incentive rather than amenities (Clark & Hunter 1992; Chen & Rosenthal 2008; Scott 2010). On the other hand, the ones of positive suggestions include, for example, studies showing that the number of hospital beds and the state of infrastructure development such as railroads correlates with the influx of population (Etzo, 2008). Similarly, a study of young people under the age of 35 from 1970 to 2000 showed that various social and environmental amenities, such as social infrastructure, influenced migration (Ito, 2005). Similarly, findings of an international study suggest that Chinese students were oriented to employment opportunities and economic well-being while Japanese students were more inclined to consider personal lifestyle and local amenities (He et al., 2016). Thus, several studies set in Japan have shown that even amenity factors could be the triggers for migration. From the above, the following hypothesis is derived.

H3: Amenity factors correlate with the population inflow rate.

*Human factors*

Population and population density have often been used as proxy variables for urbanization (e.g., Rees et al., 2017). Therefore, studies are showing that population and population density have a positive correlation with the influx population (Aoyama & Kondo, 1992; Palkama, 2018). However, some previous studies have dealt with a wider variety of human factors. Isoda (2009) and Ishiguro et al. (2012) suggest that not only economic factors, but higher education, employment, and human networks are important factors to determine migrations. In line with this, there is a study that the size of the production age population and the penetration rate of higher education positively correlate with the population influx (Palkama, 2018). Relatedly, there is evidence that the college-educated population itself becomes an amenity that attracts other well-educated migrants (Gottlieb & Joseph, 2006; Waldorf, 2007). Another study shows that the presence of some types of social capital (migration counseling and agricultural support centers) influenced the influx of population (Abe et al., 2010). From the above, the following hypothesis is derived.

H4: Human factors correlate with the population inflow rate.

*Examining a model that includes multiple factors*

Not many studies have incorporated multiple factors in the model. Most of these studies also show that economic factors influence population influx more strongly than other factors (Clark & Hunter 1992; Chen & Rosenthal 2008; Scott 2010). On the other hand, Tomioka & Sasaki (2003) targeted 208 cities in the Tohoku and Kanto regions and conducted multiple regression analysis using the inflow excess rate from 1991 to 1995 as the explained variables, and wage income per capita, land prices, minimum temperature, sewerage penetration rate, the number of beds and the number of students at universities and junior colleges as explanatory variables. As a result, it was shown that there was a significant correlation between the inflow excess rate and the minimum temperature, wage income, and land price at the 5% level, and the sewerage penetration rate, the number of beds, and the number of university / junior college students at the 10% level. However, as far as I know, there are no studies that have verified how much the analytical model that incorporates various variables is more explanatory than the model that focuses only on economic factors. On the other hand, in research in the field of human resource management, a model that includes intrinsic factors and social factors is better than a model that includes only extrinsic factors (economic factors) in explaining the organizational commitments by hierarchical multiple regression analysis (e.g., Kokubun, 2018; Kokubun & Yasui, 2020). Therefore, the following hypothesis is established.

H5: Variables composed of various factors explain population inflow better than variables composed of specific factors.

## Analytical model

In western countries, the use of gravity models has become active in population migration research since the 1960s (Greenwood and Hunt, 2003). The gravity model formulates the rule of thumb that the number of migrants between regions is proportional to the product of the population of the regions and inversely proportional to the distance between regions (Haynes & Fotheringhrum, 1984; Ravenstein, 1885). However, there are criticisms of the gravity model, which directly links migration and utility (Greenwood & Hunt, 2003). Besides, it is argued that the population size plays a large role in the gravity model, so it may not be suitable for research on economies such as Japan where the degree of concentration in the metropolitan area is strong (Ito, 2004). In my opinion, the gravity model focuses on elucidating the mechanism by which population migration occurs, but it seems that there is a lack of perspective on regional development as to what kind of land attracts the population. On the other hand, a Japanese study using the inflow rate and excess inflow rate of the population as dependent variables has shown that the inflow rate correlates with income levels (Aoyama & Kondo, 1992; Ito, 2001; Tomioka & Sasaki, 2003). Here, the inflow rate and the inflow excess rate can be calculated by the following formulas (Ito, 2001).

Inflow rate = (the number of in-migrants) / population, unit: %

Excess inflow rate = (the number of in-migrants – the number of out-migrants) / population, unit: %

Therefore, this paper does not use the gravity model but instead performs multiple regression analysis with the inflow rate and excess inflow rate of the population as the dependent variables and various factor variables as the independent variables. In this way, I would like to clarify what factors in Japan have tended to attract more population in recent years. To this end, we first look at the simple correlation between the inflow rate and excess inflow rate of the population and each variable and then perform a hierarchical multiple regression analysis with the inflow rate and excess inflow rate as the dependent variables.

## Data

The population inflow rate is the average of the values for 2010 and 2017. The value was calculated by inputting the number of Japanese in-migrants and Japanese out-migrants from other prefectures obtained from the "Basic Resident Register Population Movement Report" of the Statistics Bureau of the Ministry of Internal Affairs and Communications, and the total population (including foreigners) as of 1st October obtained from the "Population Estimate" of the Statistics Bureau of the Ministry of Internal Affairs and Communications, into the formula shown above. The reason for limiting the inflow population in the numerator to Japanese is to consider the possibility that the incentives for migration differ between Japanese and foreigners. On the other hand, the reason why foreigners are included in the denominator population is that we thought we should see how much the total local population can accept migrants from the viewpoint of regional development. Similarly, I chose October 1st instead of the beginning of the year because I thought it would be reasonable to measure capacity based on the representative population of the year. I chose the two years of 2010 and 2017 because they are relatively recent, many variables are easily available, and the researcher wanted to avoid the year that had a big impact on the domestic economy such as the Lehman shock in 2008 and the Great East Japan Earthquake in 2011.

For Gross domestic product and income per capita, the average values for 2010 and 2016 obtained from the Cabinet Office's "Prefectural Accounts" were used. For the unemployment rate, we used the average values for 2010 and 2016 obtained from the "Labor Force Survey" of the Statistics Bureau of the Ministry of Internal Affairs and Communications. For the average land price of the residential area, we used the average values for 2011 and 2017 obtained from the "Prefectural Land

Price Survey" by the Ministry of Land, Infrastructure, Transport, and Tourism. The distance from the three major metropolitan areas (Tokyo, Aichi, Osaka) is the shortest distance (geodesic length) in the spheroid (GRS80) recorded in "Distances between prefectural offices" of the Geographical Survey Institute of the Ministry of Land, Infrastructure, Transport, and Tourism. For the number of operating kilometers (km) of railway lines per 100 square kilometers, the average values for 2009 and 2013 obtained from the "Annual Report of Regional Transportation" of the Japan Transport Research Institute were used. The population density per 1 km$^2$ of the habitable land area is the average value for 2010 and 2017, which was obtained by dividing the population data recorded in the "Population Estimate" of the Statistics Bureau of the Ministry of Internal Affairs and Communications by the habitable area obtained from the "Social Life Statistics Index" of the Statistics Bureau of the Ministry of Internal Affairs and Communications.

For the ratio of people having completed up to colleges and universities, since data for multiple years was not available, only the 2010 values obtained from the "Social and Demographic System" of the Statistics Bureau of the Ministry of Internal Affairs and Communications were used. "Social capital" (Putnam, 2000), which means trust, norms, and networks, was calculated and standardized by the National Living Bureau of the Cabinet Office (2003). This value is based on the answers of 3,878 people to the 10-questionnaire survey on "friendship/exchange," "trust," and "social participation," as well as the "volunteer activity rate" and "community chest per capita" by prefecture. For the sex ratio, average age, population ratio under 15 years old, population ratio between 15 and 64 years old, and population ratio over 65 years old, the average values for 2010 and 2015 obtained from the "National Census" of the Statistics Bureau of the Ministry of Internal Affairs and Communications were used. For others, the average values for 2010 and 2017 (partly, the values of the years before and after) obtained from the "Social and Demographic System" of the Statistics Bureau of the Ministry of Internal Affairs and Communications were used.

## Result

Table 1 shows the correlation with the population inflow rate and excess inflow rate, in addition to the average value and standard deviation of each variable. From the top, economic factors, climatic factors, amenity factors, and human factors are shown in that order. Let's look at them in order.

Table 1　Mean, standard deviation, and correlation with an inflow rate

| Item | Unit | Mean | SD | Annual average population inflow rate | Annual average population inflow excess rate |
|---|---|---|---|---|---|
| Annual average population inflow rate | % | 1.568 | 0.390 | - | - |
| Annual average population inflow excess rate | % | -0.121 | 0.174 | 0.732** | - |
| *Economic factors* | | | | | |
| Gross domestic product per capita | 1,000 yen | 3768.561 | 734.762 | 0.318* | 0.512** |
| Prefectural income per person | 1,000 yen | 2777.502 | 466.708 | 0.461** | 0.642** |
| Monthly wages and salaries of household head per household | 1,000 yen | 405.687 | 43.086 | 0.396** | 0.436** |
| Cash salary (1 month per person) | 1,000 yen | 299.468 | 31.796 | 0.561** | 0.746** |
| Unemployment rate | % | 3.605 | 0.656 | 0.299* | 0.208 |
| Regional Difference Index of Consumer Prices (All items, less imputed rent) | Total = 100 | 98.876 | 1.657 | 0.474** | 0.493** |
| Average land price of residential area (per 1m$^2$) | yen | 51340.426 | 53438.780 | 0.714** | 0.751** |
| Percentage of primary industry workers | % | 12.030 | 6.658 | -0.494** | -0.703** |
| Percentage of secondary industry workers | % | 51.226 | 9.896 | -0.365* | -0.094 |
| Percentage of tertiary industry workers | % | 136.723 | 10.182 | 0.678** | 0.551** |
| Financial strength index | - | 0.503 | 0.198 | 0.602** | 0.799** |
| *Climatic factors* | | | | | |
| Yearly average of air temperature | ˚C | 15.619 | 2.307 | 0.373** | 0.246 |
| Highest temperature among monthly averages of daily highest | ˚C | 32.826 | 1.374 | 0.247 | 0.103 |
| Lowest temperature among monthly averages of daily lowest | ˚C | 0.872 | 3.162 | 0.353* | 0.252 |
| The yearly average of relative humidity | % | 69.181 | 4.390 | -0.316* | -0.377** |
| Yearly sunshine hours | hours | 1938.755 | 212.743 | 0.285 | 0.330* |
| Yearly precipitation | mm | 1748.000 | 469.135 | -0.140 | -0.098 |
| Yearly clear days | days | 24.223 | 11.798 | 0.354* | 0.320* |
| Yearly rainy days | days | 120.649 | 30.122 | -.426** | -0.264 |
| Yearly snowy days | days | 31.691 | 32.978 | -.461** | -0.344* |
| *Amenity factors* | | | | | |
| General hospitals (per 100km$^2$ of inhabitable area) | hospitals | 7.949 | 7.693 | 0.633** | 0.615** |
| General clinics (per 100km$^2$ of inhabitable area) | clinics | 112.688 | 160.298 | 0.654** | 0.654** |
| Dental clinics (per 100km$^2$ of inhabitable area) | clinics | 75.317 | 125.162 | 0.674** | 0.686** |
| Number of beds in general hospitals/clinics (per 100km$^2$ of inhabitable area) | beds | 17.336 | 17.191 | 0.660** | 0.647** |
| The diffusion rate of sewerage | % | 65.511 | 18.001 | 0.367* | 0.577** |
| Total length of roadbed (per 1km$^2$) | km | 10.343 | 9.223 | 0.677** | 0.706** |
| Total real length of roads (per 1km$^2$) | km | 4.373 | 2.304 | 0.624** | 0.735** |
| Total real length of major roads (per 1km$^2$) | km | 0.632 | 0.180 | 0.525** | 0.535** |
| The ratio of major roads paved | % | 97.745 | 1.961 | 0.424** | 0.391** |
| The ratio of local roads paved | % | 81.660 | 9.902 | 0.332* | 0.224 |
| Public parks (per 100km2 of inhabitable area) | parks | 105.617 | 122.740 | 0.683** | 0.687** |
| Persons killed or injured by traffic accidents (per 100,000 persons) | persons | 581.783 | 237.643 | 0.084 | 0.021 |
| Persons killed by traffic accidents (per 100,000 persons) | persons | 4.094 | 1.168 | -0.545** | -0.514** |
| Police men (per 1,000 persons) | persons | 1.909 | 0.309 | 0.365* | 0.249 |
| Distance from Tokyo | km | 456.651 | 322.065 | -0.132 | -0.331* |
| Distance from Aichi | km | 368.130 | 261.304 | -0.049 | -0.208 |
| Distance from Osaka | km | 367.287 | 262.428 | -0.068 | -0.109 |
| *Human factors* | | | | | |
| Population | 1,000 persons | 2710.489 | 2718.212 | 0.608** | 0.768** |
| Population density of the inhabitable area | Persons / km2 | 1364.245 | 1750.799 | 0.687** | 0.709** |
| Average age | years old | 46.556 | 1.643 | -0.578** | -0.726** |
| Population ratio under 15 years old | % | 13.072 | 0.991 | 0.098 | 0.146 |
| Population ratio 15-64 years old | % | 60.511 | 2.284 | 0.668** | 0.815** |
| Population ratio over 65 years old | % | 26.417 | 2.687 | -0.604** | -0.747** |
| The ratio of people having completed up to colleges and universities | % | 14.747 | 3.908 | 0.691** | 0.773** |
| Social capital | - | 0.000 | 0.621 | -0.211 | -0.425** |
| Population sex ratio (male per 100 females) | persons | 93.108 | 3.750 | 0.354* | 0.634** |

Note(s): n = 47; *Significance at the 5% level; **Significance at the 1% level

*Economic factors*

A significant correlation was seen in both the inflow rate and the inflow excess rate regarding gross domestic product per capita (r=0.318, p<0.05), prefectural income per person (r=0.461, p<0.01), household income (r=0.396, p<0.01), and cash salary (r=0.561, p<0.01) (the numbers in parentheses are the correlation coefficient with the population inflow rate and the p-value). On the other hand, the unemployment rate (r=0.299, p<0.05) was significantly correlated with the inflow rate, but not with the excess inflow rate. Regarding consumer prices (r=0.474, p <0.01) and average land price of residential area (r=0.714, p <0.01), there is a positive correlation with the inflow rate and the inflow excess rate. Strangely, people gather in places where commodity and land prices are high, but it makes sense to read that the causal relationship is reversed; prices rise where people gather.

Regarding the industrial composition, the percentage of primary industry workers (r=-0.494,

p <0.01), percentage of secondary industry workers (r =-0.365, p <0.05) are negative, and the percentage of tertiary industry workers (r=0.678, p<0.01) was positive, and a significant correlation was found with the inflow rate and excess inflow rate. It seems that the population inflow was smaller in prefectures where the employment scale of the tertiary industry was smaller, centered on the primary and secondary industries.

The financial strength index is the average value for the past three years obtained by dividing the standard financial income by the standard financial demand. It can be said that the higher the financial strength index, the larger the reserved financial resources for calculating the ordinary allocation tax, and the more financial resources there are (Ministry of Internal Affairs and Communications "Explanation of Indicators"). A significant correlation was shown for both the inflow rate and the inflow excess rate, and it seems that people tended to gather in prefectures with high financial strength.

Distance from Tokyo, distance from Aichi, and distance from Osaka are variables for investigating the effect of distance from the three major metropolitan areas on the inflow rate. Only distance from Tokyo shows a significant negative correlation with the excess inflow rate. This result indicates that it was difficult for people to gather as the distance from Tokyo increased. Regarding the result that the distance from Aichi and Osaka did not become significant, it may indicate the current situation that the proximity to these prefectures is becoming less likely to be an incentive for population influx as the concentration on Tokyo progresses.

From the above, correlations with the population inflow rate were found in multiple economic factor variables. This is in favor of H1.

### Climatic factors

The yearly average of air temperature and lowest temperature shows a positive, and yearly rainy days shows a negative correlation with the inflow rate. Also, the yearly average of relative humidity and yearly snowy days show a negative correlation with the inflow rate and excess inflow rate. It seems that the land with higher (minimum) temperature, less precipitation/snowfall days, and lower humidity tended to attract more people. From the above, correlations with the population inflow rate were found in multiple climate factor variables. This is in favor of H2.

### Amenity factors

The number of general hospitals, general clinics, and dental clinics, the number of beds in general hospitals/clinics and public parks (All of the above are per 100 km$^2$ of the inhabitable area), the total length of roadbed, total real length of roads, and total real length of major roads (All of the above are per 1 km$^2$), the ratio of major roads paved, and the diffusion rate of sewerage showed positive correlations with the inflow rate and excess inflow rate. Also, the ratio of local roads paved and Policemen (per 1,000 persons) show a positive correlation with the inflow rate. People seem to gather in areas with abundant amenities such as hospitals, sewers, transportation, and parks, and areas with many police officers. Also, for persons killed or injured by traffic accidents and persons killed by traffic accidents (both per 100,000 persons), the former did not show a significant correlation with the inflow rate and excess inflow rate, while the latter showed a negative correlation. This shows that it is difficult for people to gather in a land where large traffic accidents occur frequently, which causes death instead of injury. From the above, correlations with the population inflow rate were found in multiple amenity factor variables. This is in favor of H3.

*Human factors*

The ratio of people having completed up to colleges and universities, population, population density of inhabitable area, population sex ratio (male per 100 females) show a positive correlation with the inflow rate and excess inflow rate, and average age shows a negative correlation with the inflow rate and excess inflow rate. Regarding the age group, the population ratio 15-64 years old shows a positive correlation with inflow rate and excess inflow rate, while the population ratio over 65 years old shows a negative correlation with the inflow rate. It seems that the more densely populated the land is and the more college graduates, working-age population, male and young people are, the more immigrants are attracted. From the above, correlations with the population inflow rate were found in multiple human factor variables. This is in favor of H4.

Social capital has no significant correlation with the inflow rate but shows a negative correlation with the inflow excess rate. Social capital is a trust based on rural people, which can create unity and at the same time eliminate clinging and strangers (Coleman, 1988; Portes, 1998). Therefore, the negative correlation with the excess inflow rate shown here may be due to the negative nature of social capital that kept immigrants away. Alternatively, the causal relationship can be reversed. The results of several previous studies have shown that ethnic diversity brought about by population influx lowers social capital (Alesina & La Ferrara, 2002; Costa & Kahn, 2003; Putnam, 2007). The influx of strangers with different values, even if they are ethnically the same, may make it difficult for residents to unite and lower their social capital.

Table 2 Results of hierarchical multiple regression analysis

| Variable | Annual average population inflow rate | | | | | Annual average population inflow excess rate | | | | |
|---|---|---|---|---|---|---|---|---|---|---|
| *Economic factors* | | | | | | | | | | |
| Percentage of tertiary industry workers | 0.562 ** | | | | 0.614 ** | 0.367 ** | | | | |
| Cash salary (1 month per person) | 0.399 ** | | | | | 0.64 ** | | | | |
| Prefectural income per person | | | | | | | | | | 0.251 ** |
| *Climatic factors* | | | | | | | | | | |
| Yearly clear days | | | | | 0.240 ** | | | | | |
| Yearly snowy days | | -0.461 ** | | | | | | | | |
| The yearly average of relative humidity | | | | | | | -0.377 ** | | | |
| *Amenity factors* | | | | | | | | | | |
| General clinics (per 100km2 of the inhabitable area) | | | 0.604 ** | | | | | | | |
| The ratio of major roads paved | | | 0.335 ** | | | | | 0.247 ** | | |
| The total real length of roads | | | | | | | | 0.521 ** | | 0.263 ** |
| The total length of roadbed (per 1km2) | | | | | 0.334 ** | | | | | 0.184 * |
| The diffusion rate of sewerage | | | | | | | | 0.374 ** | | |
| *Human factors* | | | | | | | | | | |
| Population density of the inhabitable area | | | | 0.352 * | | | | | | |
| Population ratio 15-64 years old | | | | 0.368 * | | | | | 0.537 ** | 0.505 ** |
| The ratio of people having completed up to colleges and universities | | | | 0.346 * | | | | | 0.392 ** | |
| Social capital | | | | 0.339 ** | 0.269 ** | | | | | |
| Population sex ratio (male per 100 females) | | | | | 0.261 ** | | | | | |
| R² | 0.605 | 0.213 | 0.537 | 0.655 | 0.762 | 0.680 | 0.142 | 0.695 | 0.741 | 0.811 |
| Adjusted R² | 0.587 | 0.195 | 0.516 | 0.622 | 0.733 | 0.665 | 0.123 | 0.674 | 0.729 | 0.793 |
| F | 33.673 ** | 12.168 ** | 25.563 ** | 19.948 ** | 26.289 ** | 46.659 ** | 7.445 ** | 32.645 ** | 62.978 ** | 45.064 ** |

Note(s): n = 47; *Significance at the 5% level; **Significance at the 1% level

*Multiple regression analysis*

Table 2 is the result of multiple regression analysis. The left half is the inflow rate, and the right half is the analysis result with the inflow excess rate as the dependent variable. From the left, the variables were input for each category in the order of economic factors, climatic factors, amenities, and human factors, and finally, all variables were input. Since there are many variables, the variable selection was performed by the stepwise method. Looking at the analysis results with the inflow rate as the dependent variable, as for the economic factors, percentage of tertiary industry workers ($\beta$=0.562, $p$<0.01) and cash salary (1 month per person) ($\beta$=0.399, $p$<0.01) were selected as significant variables showing a positive correlation with migration rate. Next, as for the climate variables, yearly snowy days ($\beta$= -0.461, $p$<0.01) showed a negative and significant correlation. As for amenity variables, general clinics (per 100 km$^2$ of the inhabitable area) ($\beta$=0.604, $p$<0.01) and the ratio of major roads paved ($\beta$=0.335, $p$<0.01) showed positive and significant correlations. These are variables that were significant even with a simple correlation.

On the other hand, as for human factors, in addition to the population density of inhabitable area ($\beta$=0.352, $p$<0.05), population ratio 15-64 years old ($\beta$=0.368, $p$<0.05), the ratio of people having completed up to colleges and universities ($\beta$=0.346, $p$<0.05), social capital ($\beta$=0.339, $p$<0.01) showed a positive correlation. Since social capital did not show a significant correlation by simple correlation, it seems that the correlation was shown by controlling the influence of other variables in multiple regression analysis. In other words, it seems that prefectures with a large number of population per land, university graduates, and working-age population, and with abundant social capital tended to attract more people. Finally, when all variables were input, percentage of tertiary industry workers ($\beta$=0.614, $p$<0.01) of the economic factors, yearly clear days ($\beta$=0.240, $p$<0.01) of the climatic factors, the total length of roadbed ($\beta$ = 0.334, $p$ <0.01) of the amenity factors, and social capital ($\beta$=0.269, $p$<0.01) and population sex ratio ($\beta$=0.261, $p$ <0.01) of the human factors were selected.

Comparing the adjusted R-squared, the model in which all variables were input was 0.762, whereas the one for the economic factor only 0.605, the climatic factor only 0.213, the amenity factor only 0.537, and the human factor only 0.655. That is, the adjusted R-squared increases in the range of 0.111 to 0.538 in the model in which all variables are input, as compared with the case where variables of only individual categories are used. According to Cohen's (1988) criteria, 0.02 is small, 0.13 is medium, and 0.26 is large, which means that the model has improved at least to a degree close to "medium". This is in favor of H5. Even in the case with the excess inflow rate as the dependent variable, the model in which all variables are input is better in the adjusted R-squared than the model in which each factor is input individually, although the composition of the selected variables is different. However, the adjusted R-squared of the model with only human factors is 0.729, and the difference from the adjusted R-squared of the model using all variables 0.793 is 0.064 close to "small".

The Average land price of residential area is excluded from the variables of multiple regression analysis because the correlation coefficient with the inflow rate and the excess inflow rate both exceeds 0.7 and the problem of multicollinearity is suspected. By the way, when this variable is included, it is selected as an independent variable in the inflow rate model, and the total length of roadbed (per 1 km$^2$) and population sex ratio (male per 100 females) are not selected instead. However, adjusted R-squared decreased from 0.733 to 0.616, indicating that including the average land price of the residential area as an independent variable reduces the explanatory power of the model. On the other hand, it was not selected as a variable in the model with the excess inflow rate as the dependent variable.

Figure 1 is a scatter plot of the population inflow rate, with the predicted value calculated from the regression equation in Table 2 on the horizontal axis and the measured value on the vertical axis. This means that prefectures above the diagonal are attracting more influx than expected, and prefectures below the diagonal are attracting less influx than expected. It seems that such a difference is caused by another factor that was not included in the model estimated in this paper.

## Discussion

We analyzed to clarify what factors in a prefecture collect the population influx, using population inflow rate and excess inflow rate as dependent variables, and various variables related to economic factors, climatic factors, amenity factors, and human factors as independent variables. Also, we analyzed to clarify how much the explanatory power of the model using various elements is improved compared to the model using individual elements. First, as a result of simple correlation analysis, it was shown that many variables have correlations with the inflow rate and the inflow excess rate. Next, as a result of the hierarchical multiple regression analysis by the stepwise method, the inflow rate was more correlated with independent variables when all the four-factor variables were used compared with the case where only the variables of a single factor, for example, the economic factor are used. It was shown that the model with the inflow rate as a dependent variable showed a moderate improvement in explanatory power, and the model with the excess inflow rate as a dependent variable showed a small improvement in explanatory power, in the multiple factor model compared with the single-factor model. In other words, in addition to excellent economic factors such as a high proportion of tertiary industries and high levels of salary and income, the results show that comfortable climatic factors including many sunny days, few snowy days, and low humidity, sufficient amenity factors such as hospitals, railroads, roads, and sewage, abundant human factors including dense population, a large volume of the working-age population, men, and college graduates, and rich social capital are important for attracting the population.

Figure 1 Scatter plot of predicted and actual population inflow rates

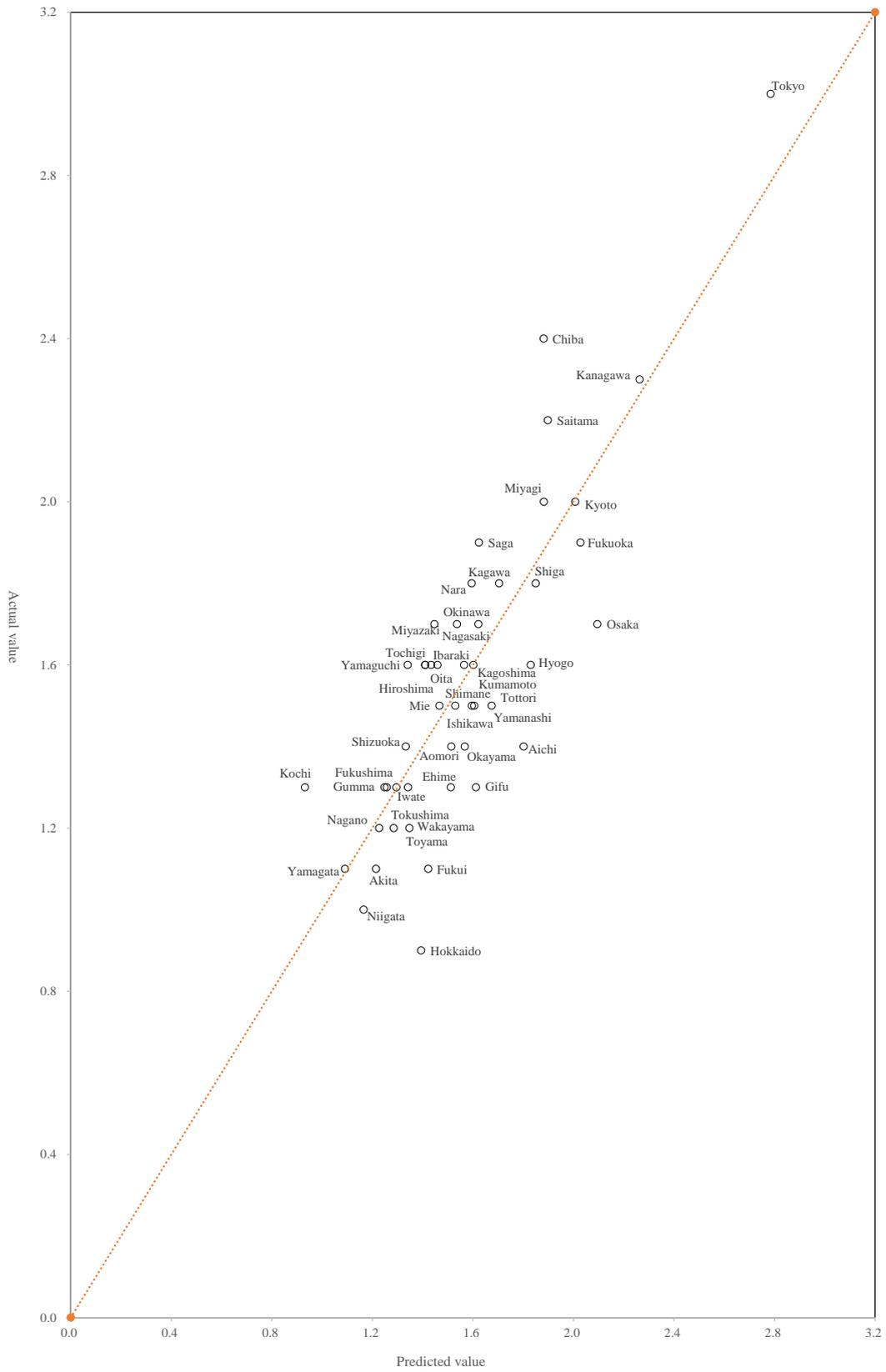

Comparing the inflow rate model and the inflow excess rate model, the composition of the independent variables is slightly different. Therefore, in the model in which the excess inflow rate is used as the dependent variable, the difference in the adjusted R-squared is relatively small between the case where only the variables of the human factor are used and the case where all the variables are used. It can be said that most of the excess inflow rate can be explained by only two human-related variables, the working-age population ratio, and the university graduate ratio. By the way, the adjusted R-squared of the model using these two variables is even larger than the model using the two economic factor variables of the tertiary industry ratio and the amount of salary by Cohen's standard "small" or more. Therefore, in addition to being able to explain the population inflow rate better with a model composed of various variables than with a model composed of variables with a specific element, human factor variables explain the population inflow rate better than economic factor variables. This is a new finding contributing to the literature.

Why is the explanatory power of economic factor variables smaller than the explanatory power of human factor variables? The most straightforward view of the analysis would be that human factor variables can better explain population migration than economic factor variables. But here I would like to set aside this possibility and consider another reason because moving land in search of high salaries is the most economically natural phenomenon. So, since Hicks (1932), many previous studies have focused on economic factors in the analysis of population movement. One possible cause is the difference in position between economic agents. For example, high wage levels are an incentive for workers, but a disadvantage for employers. In this way, there is a large difference in how economic factor variables are perceived depending on the migrant's position, so it is considered that the difference in reactivity between the two was offset in the analysis of this paper for the migration of diverse people, and the correlation was reduced accordingly.

Another possibility is the difference between generations. Previous studies have shown that young people tend to move to higher-income prefectures and older people tend to move to lower-income prefectures. The authors of the paper speculate that many of the former were retired people and that the emphasis was on livability rather than income (Tamura & Sakamoto, 2016). Interestingly, Tamura & Sakamoto's research shows that both young and old people tend to move to prefectures with many young people. This is consistent with the results of this paper, which show that human factors such as the working-age population have high explanatory power.

Social capital showed a simple negative correlation with the excess inflow rate. However, multiple regression analysis controlling other variables showed a positive correlation with the inflow rate. How can this result be interpreted? Social capital creates trust and unity among residents and sometimes leads to exclusion of strangers (Coleman, 1988; Portes, 1998; Takagishi & Kinan, 2012). However, some studies have shown that a positive aspect of social capital, high trust, leads to a positive attitude towards influxes by increasing tolerance to uncertainty (Economidou et. al., 2020; Herreros

& Criado, 2009; Rustenbach, 2010)). In Japan as well, in addition to measures such as subsidies, it has been shown that measures that utilize the social capital of residents play an important role in promoting migration and settlement (Sakuno, 2016; Takeda & Kaga,). 2018). Therefore, the results of multiple regression analysis would mean that social capital alone is weak and could be sometimes harmful as a factor to attract people, but in a land where other factors are prepared, it helps attract people by making good use of it. Recent studies have shown that social capital is effective in efforts to prevent Covid-19 infection (e.g., Kokubun, 2020; Kokubun et al., 2020), so it can be said that it is becoming a factor that cannot be ignored when considering population movement in the with-corona era.

## Implication

Today, in response to the corona pandemic, it is expected that migration to rural areas, correction of overconcentration in large cities, and regional promotion will progress. The results of this paper show what kind of factors have attracted the population to a prefecture in recent years, and can be used as a reference in promoting the flow of migration to rural areas. In particular, the result that the correlation between the economic factor variables only and population inflow is smaller than the correlation between the variables of diverse elements including climatic, amenity, and human factors in addition to economic factors and population inflow will be good news for considering promotion measures in lacking areas. In particular, it will be important to consider how to improve human factors retaining university graduates of working age in the region that have a large association with the excess inflow rate. At the same time, social capital, which was correlated with the inflow rate, is one of the few resources that these regions have, and therefore strategies such as utilizing it to support immigrants physically and mentally will be realistic and successful.

## Limitation

The analysis results in this paper are based on cross-sectional analysis and do not show a causal relationship. Also, since data for a specific year is used, there remains anxiety in terms of robustness. Furthermore, since it uses data before the corona pandemic, it is unclear how much it applies to the era of with-corona. In the future, It is preferable to verify the results of this paper by analyzing to clarify what kind of difference has occurred in the population inflow between prefectures that have taken specific measures and those that have not taken, and what kind of correlation there is between variables depending on the time of analysis. Furthermore, it seems possible to supplement the limitation of this paper by considering and analyzing newly needed factors in the era of with-corona.

## Conclusion

Regional promotion and centralized correction in Tokyo have long been the goals of the Government of Japan. Furthermore, in the wake of the recent new coronavirus (COVID-19) epidemic, the momentum for rural migration is increasing, to prevent the risk of infection with the help of penetration of remote work. However, there is not enough debate about what kind of land will attract the population. Therefore, in this paper, we will consider this problem by performing correlation analysis and multiple regression analysis with the inflow rate and the excess inflow rate of the population as the dependent variables, using recent government statistics for each prefecture. As a result of the analysis, in addition to economic factor variables, variables of climatic, amenity, and human factors correlated with the inflow rate, and it was shown that the model has the greatest explanatory power when multiple factors were used in addition to specific factors. Therefore, local prefectures are required to take regional promotion measures focusing on not only economic factors but also multifaceted factors to attract the outside population.

## References

Abe, S., Kondo, A., and Kondo, A. (2010) Factors analysis of "UIJ-turn" migration and policies of population inflow. *Infrastructure Planning Review*, 27, 219-230. Retrieved from http://library.jsce.or.jp/jsce/open/00041/2010/27-0219.pdf

Aoyama, Y. & Kondo. A. (1992) A Migration Model Based on the Difference in Utility between Regions. *Infrastructure Planning Review*, 10, 151-158. https://doi.org/10.2208/journalip.10.151

Alesina, A., & La Ferrara, E. (2002). Who trusts others?. *Journal of Public Economics*, 85(2), 207-234. https://doi.org/10.1016/S0047-2727(01)00084-6

Arntz, M. (2010). What attracts human capital? Understanding the skill composition of interregional job matches in Germany. *Regional Studies*, 44(4), 423-441. https://doi.org/10.1080/00343400802663532

Asada, Y. (1996). Interregional migration in the postwar Japan: economic analysis using regional income data. *Departmental Bulletin Paper, Osaka Prefecture University*, 41(2), 91-125. http://doi.org/10.24729/00001452

Attanasio, O. & Padoa-Schioppa, F. (1991). Regional Inequalities, Migration and Mismatch in Italy, 1960–1986. In F. Padoa-Schioppa (ed.), *Mismatch and Labour Mobility*, Cambridge, Cambridge University Press.

Cabinet Office. *Prefectural accounts*. Retrieved from https://www.esri.cao.go.jp/jp/sna/data/data_list/kenmin/files/contents/main_h28.html (Accessed: September 6, 2020)

Cabinet Office National Living Bureau (2003). *Sōsharu kyapitaru: Yutakana ningen kankei to shimin katsudō no kō junkan o motomete (Social Capital: Seeking a virtuous cycle of rich relationships* *and civic activities)*. (in Japanese)  https://www.npo-homepage.go.jp/toukei/2009izen-chousa/2009izen-sonota/2002social-capital (Accessed: September 6, 2020)

Chen, Y., & Rosenthal, S. S. (2008). Local amenities and life-cycle migration: Do people move for jobs or fun?. *Journal of Urban Economics*, 64(3), 519-537. https://doi.org/10.1016/j.jue.2008.05.005

Clark, D. E., & Hunter, W. J. (1992). The impact of economic opportunity, amenities and fiscal factors on age-specific migration rates. *Journal of Regional Science*, 32(3), 349-365. https://doi.org/10.1111/j.1467-9787.1992.tb00191.x

Cohen, J. (1988). *Statistical power analysis for the behavioral sciences* (2nd ed.). Hillsdale, NJ: Erlbaum.

Coleman, J. S. (1988). Social capital in the creation of human capital. *American Journal of Sociology*, 94, S95-S120. https://doi.org/10.1086/228943

Costa, D. L., & Kahn, M. E. (2003). Civic engagement and community heterogeneity: An economist's perspective. *Perspectives on Politics*, 1(1), 103-111. https://doi.org/10.1017/S1537592703000082

Economidou, C., Karamanis, D., Kechrinioti, A., & Xesfingi, S. (2020). The Role of Social Capital in Shaping Europeans' Immigration Sentiments. *IZA Journal of Development and Migration*, 11(1), 20200003, eISSN 2520-1786. https://doi.org/10.2478/izajodm-2020-0003

Etzo, I. (2008). Determinants of interregional migration in Italy: A panel data analysis. http://dx.doi.org/10.2139/ssrn.1135165

Faggian, A., McCann, P., & Sheppard, S. (2007). Human capital, higher education and graduate migration: an analysis of Scottish and Welsh students. *Urban Studies*, 44(13), 2511-2528. https://doi.org/10.1080/00420980701667177

Ferguson, M., Ali, K., Olfert, M. R., & Partridge, M. (2007). Voting with their feet: jobs versus amenities. *Growth and Change*, 38(1), 77-110. https://doi.org/10.1111/j.1468-2257.2007.00354.x

Gottlieb, P. D., & Joseph, G. (2006). College-to-work migration of technology graduates and holders of doctorates within the United States. *Journal of Regional Science*, 46(4), 627-659. https://doi.org/10.1111/j.1467-9787.2006.00471.x

Graves, P. E. (1980). Migration and climate. *Journal of Regional Science*, 20(2), 227-237. https://doi.org/10.1111/j.1467-9787.1980.tb00641.x

Greenwood, M. J., & Hunt, G. L. (2003). The early history of migration research. International *Regional Science Review*, 26(1), 3-37. https://doi.org/10.1177/0160017602238983

Haynes, K., & Fotheringhrum, A. (1984). Gravity and spatial interaction models. Beverly Hills, Sage.

He, Z., Zhai, G., Asami, Y., & Tsuchida, S. (2016). Migration intentions and their determinants: Comparison of college students in China and Japan. *Asian and Pacific Migration Journal*, 25(1), 62-84. https://doi.org/10.1177/0117196815621203

Herreros, F., & Criado, H. (2009). Social trust, social capital and perceptions of immigration. *Political Studies*, 57(2), 337-355. https://doi.org/10.1111/j.1467-9248.2008.00738.x

Herzog, H. W., & Schlottmann, A. M. (1986). The metro rating game: what can be learned from the recent migrants?. *Growth and Change*, 17(1), 37-50. https://doi.org/10.1111/j.1468-2257.1986.tb00931.x

Hicks, J. (1932). *The Theory of Wages*, London, Macmillan.

Higuchi, Y. (1991). *Nihon Keizai to Shugyo Kodo (Japanese Economy and Job Search Behavior)*, Tokyo, Toyo Keizai. (in Japanese)

Hirata, S., Kawabata, Y., & Fujii, S. (2019). A Study on the effect of investment in road infrastructures on the population concentration into Tokyo. *Journal of Japan Society of Civil Engineers*, 75(5), 967-978. Retrieved from https://www.jstage.jst.go.jp/article/jscejipm/75/5/75_I_967/_article/-char/ja

Ishikawa, Y. (2020). Internal Migration in Japan. In Bell, M., Bernard, A., Charles-Edwards, E., Zhu, Y. (eds), *Internal Migration in the Countries of Asia*, 113-136, Cham, Springer.

Isoda, N. (2009). Higher education and population concentration into Tokyo metropolitan area in Japan. *Fukuoka University Review of Literature & Humanities*, 41(3), 1029 - 1052. Retrieved from https://fukuoka-u.repo.nii.ac.jp/index.php?active_action=repository_view_main_item_detail&page_id=13&block_id=39&item_id=1038&item_no=1

Ito, K. (2001). Sengonihon no jinkō idō ni taisuru shotokukakusa-setsu no setsumei-ryoku to kongo no kadai (Explanatory power of income disparity theory for postwar Japan's migration and future issues). *Journal of Region and Society*, 4, 9-38. (in Japanese) Retrieved from https://core.ac.uk/download/pdf/233904961.pdf

Ito, K. (2004). Kokunai chōkyori jinkō idō ni ataeru seikatsu suijun no eikyō ni tsuite - shin kokumin seikatsu shihyō to 1990-nen kokusei chōsa shūkei kekka o riyō shite (Impact of living standards on domestic long-distance migration-using the New National Living Index and the results of the 1990 National Survey). *Review of Economics and Information Study*, 4(1-4), 662-692. (in Japanese)

Ito, K. (2006). An analysis of income growth effect on the long-distance migration in postwar Japan. *Journal of Population Studies*, 38, 89-98. https://doi.org/10.24454/jps.38.0_89

Ito, K. (2008). Internal long-distance migration in Japan: comparison of the methods of the analysis on the determinants and some notes. *Review of Economics and Information Studies*, 8(3-4), 31-64.

Japan Transport and Tourism Research Institute. *Chiiki Kōtsū Nenpō (Regional Transportation Annual Report)*. (in Japanese)

Kokubun, K. (2018). Education, organizational commitment, and rewards within Japanese manufacturing companies in China. *Employee Relations* 40(3), 458-485. https://doi.org/10.1108/ER-12-2016-0246

Kokubun, K. (2020). Social capital may mediate the relationship between social distance and COVID-19 prevalence. *arXiv preprint* arXiv:2007.09939. Retrieved from https://arxiv.org/abs/2007.09939

Kokubun, K., Ino, Y., & Ishimura, K. (2020). Social capital and resilience make an employee cooperate for coronavirus measures and lower his/her turnover intention. *arXiv preprint* arXiv:2007.07963. Retrieved from https://arxiv.org/abs/2007.07963 Accessed 8 Aug 2020

Kokubun, K., & Yasui, M. (2020). The difference and similarity of the organizational commitment–rewards relationship among ethnic groups within Japanese manufacturing companies in Malaysia. *International Journal of Sociology and Social Policy*. (ahead-of-print) https://doi.org/10.1108/IJSSP-03-2020-0099

Lee, Y. (2012). Economic gains youth receive from inter- regional migration. In Ishiguro, I. et al. (eds), *Tokyoni Deru Wakamonotachi (The Brain Drain: Why Japanese Youth Move to Tokyo)*, Minerva Shobo, 47- 90. (in Japanese)

Lee, Y. J., & Sugiura, H. (2018). Key Factors in Determining Internal Migration to Rural Areas and Its Promoting Measures? A Case Study of Hirosaki City, Aomori Prefecture. *Public Policy Review*, 14(1), 153-176. Retrieved from https://www.mof.go.jp/english/pri/publication/pp_review/fy2017/ppr14_01_06.pdf

Liu, Y., & Shen, J. (2014). Spatial patterns and determinants of skilled internal migration in China, 2000–2005. *Papers in Regional Science*, 93(4), 749-771. https://doi.org/10.1111/pirs.12014

Ministry of Health, Labor, and Welfare. *Iryō shisetsu chōsa (Medical facility survey)*. (in Japanese) Retrieved from https://www.mhlw.go.jp/toukei/list/79-1.html (Accessed: September 6, 2020)

Ministry of Internal Affairs and Communications. Explanation of indicators. Retrieved from https://www.soumu.go.jp//000264701.pdf (Accessed: September 6, 2020)

Ministry of Land, Infrastructure, Transport, and Tourism. *Todōfuken chika chōsa (Prefectural land price survey)*. (in Japanese) Retrieved from https://www.mlit.go.jp/totikensangyo/totikensangyo_fr4_000044.html (Accessed: September 6, 2020)

Geographical Survey Institute, Ministry of Land, Infrastructure, Transport and Tourism. *Todōfuken-chō-kan no kyori (Distance between prefectures)*. (in Japanese) Retrieved from https://www.gsi.go.jp/KOKUJYOHO/kenchokan.html (Accessed: September 6, 2020)

Ohta, S. & Y. Ohkusa. (1996). Regional Labor Mobility and Wage Curve in Japan. *JCER Economic Journal*, 32, 111-132.

Palkama, J. (2018). *The determinants of internal migration in Finland*. Retrieved from https://aaltodoc.aalto.fi/handle/123456789/35572

Piras, R. (2012). Internal migration across Italian regions: macroeconomic determinants and accommodating potential for a dualistic economy. *The Manchester School*, 80(4), 499-524. https://doi.org/10.1111/j.1467-9957.2011.02278.x

Porter, M. (1990). *The Competitive Advantage of Nations*, New York, Free Press.

Putnam, R. D. (2000). *Bowling Alone: The Collapse and Revival of American Community*. New York, Simon & Schuster.

Putnam, R. D. (2007). E pluribus unum: Diversity and community in the twenty-first century the 2006 Johan Skytte Prize Lecture. *Scandinavian Political Studies*, 30(2), 137-174. https://doi.org/10.1111/j.1467-9477.2007.00176.x

Portes, A. (1998). Social capital: Its origins and applications in modern sociology. *Annual Review of Sociology*, 24, 1-24. https://doi.org/10.1146/annurev.soc.24.1.1

Ravenstein, E. G. (1885). *The laws of migration, Journal of Royal Statistical Society*, 48, 167 - 235. https://doi.org/10.2307/2979333

Rees, P., Bell, M., Kupiszewski, M., Kupiszewska, D., Ueffing, P., Bernard, A., Charles-Edwards, E., & Stillwell, J. (2017). The impact of internal migration on population redistribution: An international comparison. *Population, Space and Place*, 23(6), e2036. https://doi.org/10.1002/psp.2036

Rustenbach, E. (2010). Sources of negative attitudes toward immigrants in Europe: A multi-level analysis. *International Migration Review*, 44(1), 53-77. https://doi.org/10.1111/j.1747-7379.2009.00798.x

Sakuno, H. (2016). The Increase of Migrants into Local Areas and Regional Correspondence: What does "Return to the Country" Mean for Local Areas? *Annals of the Japan Association of Economic Geographers*, 62(4), 324-345. Retrieved from https://www.jstage.jst.go.jp/article/jaeg/62/4/62_324/_pdf

Scott, A. J. (2010). Jobs or amenities? Destination choices of migrant engineers in the USA. *Papers in Regional Science*, 89(1), 43-63. https://doi.org/10.1111/j.1435-5957.2009.00263.x

Statistics Bureau, Ministry of Internal Affairs and Communications. *Jinkō suikei (Population estimation)*    Retrieved from http://www.stat.go.jp/data/jinsui/index2.html

Statistics Bureau, Ministry of Internal Affairs and Communications. *Social life statistical index*. Retrieved from https://www.stat.go.jp/data/shihyou/naiyou.html    (Accessed: September 6, 2020)

Statistics Bureau, Ministry of Internal Affairs and Communications. *Jūmin kihon daichō jinkō idō hōkoku (Basic resident register population migration report)* Retrieved from https://www.e-stat.go.jp/stat-search/files?page=1&layout=datalist&toukei=00200523&tstat=000000070001&cycle=7&year=20190&month=0&tclass1=000001011680    (Accessed: September 6, 2020)

Statistics Bureau, Ministry of Internal Affairs and Communications. *Census*. Retrieved from https://www.e-stat.go.jp/stat-search/files?page=1&layout=datalist&toukei=00200521&tstat=000001049104&cycle=0&tclass1=000001049105 (Accessed: September 6, 2020)

Statistics Bureau, Ministry of Internal Affairs and Communications. *Labor Force Survey*. Retrieved from https://www.e-stat.go.jp/stat-search/files?page=1&layout=datalist&toukei=00200531&tstat=000001110001&cycle=0&tclass1=000001011635&tclass2=000001011637 (Accessed: September 6, 2020)

Statistics Bureau, Ministry of Internal Affairs and Communications. *Social / demographic system*. Retrieved from https://www.e-stat.go.jp/stat-search/files?page=1&layout=datalist&toukei=00200502&tstat=000001137306&cycle=0&tclass1=000001137307&result_page=1 (Accessed: September 6, 2020)

Storper, M., & Scott, A. J. (2009). Rethinking human capital, creativity and urban growth. *Journal of Economic Geography*, 9(2), 147-167. https://doi.org/10.1093/jeg/lbn052

Tachi, M. (1963). Shotoku no chiiki bunseki to kokunai jinkō idō: Demogurafi no kenchi kara, Guranto shohan hakkō san hyaku-nen o kinen shite (Regional analysis of income and domestic migration: From a demographic point of view, to commemorate the 300th anniversary of the publication of Grant's first edition). *Hitotsubashi University Research Series. Economics*, 7, 179-246. (in Japanese) http://doi.org/10.15057/9383

Tanioka, K. (2001). Study on regional income disparity and population migration. *Journal of Region and Society*, 4, 58. (in Japanese) Retrieved from Retrieved from https://core.ac.uk/download/pdf/233904962.pdf

Takagishi, M., & Kiminami, L. (2012). I-Turn Promotion in Rural Areas: Case Study from Chichibu City, Saitama Prefecture. *Bulletin of the Faculty of Agriculture, Niigata University*, 65(1), 1-14. Retrieved from https://agriknowledge.affrc.go.jp/RN/2010834121.pdf

Takeda, Y., & Kaga, A. (2018). A study of the policy at migration and settlement in regional hub cities and dweller's characteristics. *Journal of the City Planning Insttute of Japan*, 53(3), 1153-1160. https://doi.org/10.11361/journalcpij.53.1153

Tamura, K. & Sakamoto, H. (2016). Nihon no todōfuken-kan jinkō idō no sedai-kan hikaku (Intergenerational comparison of Japan's inter-prefectural migration). *AGI Working Paper Series*, 2016, 1-11. (in Japanese) Retrieved from http://id.nii.ac.jp/1270/00000114/

Tomioka, T. & Sasaki, K. (2003). Jinkō idō o kōryo shita toshi ameniti no keizai-teki hyōka (Economic evaluation of urban amenities considering population migration). *Journal of Applied Regional Science*, 8(2), 33-44.

Toyoda, T. (2013). Changes in regional income inequality and migration in Japan: Using estimated household income adjusted for household size and age compositions. *Annals of the Association *of Economic Geographers*, 59(1), 4-26.   https://doi.org/10.20592/jaeg.59.1_4

Vakulenko, E. S. (2016). Econometric analysis of factors of internal migration in Russia. *Regional Research of Russia*, 6(4), 344-356. https://doi.org/10.1134/S2079970516040134

Waldorf, B. S. (2007). Brain drain in rural America. *Selected Paper prepared for presentation at the American Agricultural Economics Association Annual Meeting*, Portland, Oregon, July 28-30, 2007.

Watanabe, M. (1994). *Chiiki keizai to jinkō (Regional economy and population)*, Tokyo, Nippon Hyoron Sha.

Whisler, R. L., Waldorf, B. S., Mulligan, G. F., & Plane, D. A. (2008). Quality of life and the migration of the college-educated: a life-course approach. *Growth and Change*, 39(1), 58-94. https://doi.org/10.1111/j.1468-2257.2007.00405.x

Zhang, J., Seya, H., Kaneshige, H., & Chikaraishi, M. Longitudinal analysis of factors affecting inter-prefecture population mobility based on a discrete choice model with spatial context dependency. *Geographical Sciences*, 71(3), 118-132. https://doi.org/10.20630/chirikagaku.71.3_118